\documentclass[aps,prb,final,twocolumn,superscriptaddress]{revtex4-1}

\usepackage{graphicx}
\usepackage{epsfig}
\usepackage{natbib}
\usepackage{amsmath}

\newcommand{\CrNbS}{$\mathrm{CrNb}_3\mathrm{S}_6$}

\begin{document}


\title{Incommensurate--commensurate transitions in the mono-axial 
chiral helimagnet driven by the magnetic field}


\author{Victor Laliena}
\email[]{laliena@unizar.es}

\author{Javier Campo}
\email[]{javier.campo@csic.es}
\affiliation{Instituto de Ciencia de Materiales de Arag\'on (CSIC -- University of Zaragoza), 
C/Pedro Cerbuna 12, 50009 Zaragoza, Spain}

\author{Jun-Ichiro Kishine}
\affiliation{Division of Natural and Environmental Sciences The Open University of Japan, 
Chiba, 261-8586, Japan}
\affiliation{Centre for Chiral Science, Hiroshima University, Higashi-Hiroshima, 
Hiroshima 739-8526, Japan}

\author{Alexander S. Ovchinnikov}
\affiliation{Institute of Natural Sciences, Ural Federal University, 
Ekaterinburg, 620083, Russia}

\author{Yoshihiko Togawa}
\affiliation{N2RC, Osaka Prefecture University, 1-2 Gakuencho, Sakai, Osaka 599-8570, Japan}
\affiliation{Centre for Chiral Science, Hiroshima University, Higashi-Hiroshima, 
Hiroshima 739-8526, Japan}

\author{Yusuke Kousaka}
\author{Katsuya Inoue}
\affiliation{Department of Chemistry, Faculty of Science, Hiroshima University, 
Higashi-Hiroshima, Hiroshima 739-8526, Japan}
\affiliation{Centre for Chiral Science, Hiroshima University, Higashi-Hiroshima, 
Hiroshima 739-8526, Japan}


\date{February 1, 2016}

\begin{abstract}
The zero temperature phase diagram of the mono-axial chiral helimagnet in the magnetic field plane
formed by the components parallel and perpendicular to the helical axis is thoroughly analyzed. 
The nature of the transition to the commensurate state depends on the angle between the field 
and the helical axis. For field directions close to the directions parallel or perpendicular 
to the helical axis the transition 
is continuous, while for intermediate angles the transition is discontinuous and the incommensurate 
and commensurate states coexist on the transition line. The continuous and discontinuous
transition lines are separated by two tricritical points with specific singular behaviour.
The location of the continuous and discontinuous lines and of the tricritical 
points depend strongly on the easy-plane anisotropy, the effect of which is analyzed.
For large anisotropy the conical approximation locates the transition line very accurately, although
it does not predict the continuous transitions nor the tricitical behaviour. It is shown that for 
large anisotropy, as in \CrNbS, the form of the transition line is universal, that is,
independent of the sample, and obeys a simple equation. The position of the tricritical points, 
which is not universal, is theoretically estimated for a sample of \CrNbS. 
\end{abstract}

\pacs{111222-k}
\keywords{Helimagnet, Dzyaloshinskii--Moriya interaction, Chiral Soliton Lattice}

\maketitle


\section{Introduction}

Chiral magnets are very promising ingredients for spintronic based devices since they support 
peculiar magnetic textures that affect the charge and spin transport properties in different 
ways \cite{Kishine15}. As these magnetic textures can be deeply altered by magnetic fields,
the transport properties can be magnetically controlled, and thus they might be used as 
functional components of magnetic devices\cite{Fert13,Romming13}.

For example, in the mono-axial chiral 
helimagnet, an instance of which is \CrNbS, the competition between
the ferromagnetic (FM) and Dzyaloshinskii--Moriya (DM) interactions leads to an incommensurate 
magnetic helix propagating with period $L_0$ along a crystallographic axis, which 
will be called here the DM axis. 
The effect of the magnetic field on the helical magnetic ordering is an old topic and was discussed 
for instance in Ref.~\onlinecite{Nagamiya68}. In the case of mono-axial DM interactions,
application of a magnetic field deforms the helix in a way which depends on the field direction. 
If the field is parallel to the DM axis, a conical helix is 
formed, with the spins tilted by a constant angle towards the DM axis, while revolving round it
with a period $L_0$ independent of the magnetic field. 
If, on the other hand, the magnetic field is perpendicular to the DM axis, the spins remain 
perpendicular to the DM axis but tend to be aligned with the magnetic field. 
The spins rotate slowly about the DM axis in the regions where they are nearly parallel to the 
magnetic field, and the rotation becomes faster as the spin direction separates from the field 
direction. As the field increases, the regions where the spins rotate slowly become very wide 
and the regions where the spins rotate fast become very narrow, and a Chiral Soliton Lattice 
(CSL) is formed \cite{Kishine15,Dzyal58,Dzyal64,Izyumov84}.
This CSL, which is realised in \CrNbS\ \cite{Togawa12}, supports dynamic modes like coherent sliding 
\cite{Kishine12} and gives rise to exciting phenomena as spin motive forces \cite{Ovchinnikov13} and 
tuneable magnetoresistence \cite{Kishine11,Togawa13,Ghimire13,Bornstein15}.

If the applied field is strong enough, the incommensurate chiral helicoid is transformed in a 
commensurate FM state, called the forced ferromagnetic (FFM) state. 
Chiral and incommensurate (IC) to commensurate (C) transitions, 
which are universal and appear in many branches of physics, take place in these systems.
The nature of the IC-C transition and its driving mechanism depend on the angle $\alpha$ 
between the field and the DM axis. If a field parallel to the DM axis increases, the cone on which
the spins lie becomes narrower, but the helix period $L_0$ remains constant.
At a certain critical field the cone closes completely and the FFM state is reached 
continuously.  On the other hand, the effect of increasing a field perpendicular to the DM axis 
is to increase the period of the CSL, so that the FFM state is reached continuously
at a critical field at which the period $L$ of the CSL diverges \cite{Kishine15}. Thus, in the two 
limiting cases of magnetic field parallel and perpendicular to the DM axis the FFM state is reached 
\textit{via} two very different mechanisms. On the other hand, even if these systems have been 
extensively studied, the nature of the transition to the FFM has not been elucidated yet when 
$\alpha$ is continuously varied between the limiting cases $0$ and $\pi/2$. 

A theoretical attempt has been made recently based on an approximation which makes the model 
equations easily solvable \cite{Chapman14}. As we shall show in section~\ref{sec:approx} the
approximation gives the transition line very accurately when the easy-plane anisotropy is large,
as in the case of \CrNbS. However, the nature of the transition, and in particular its continuity
or discontinuity, is not well described within this approximation.
In this work we address the important question of the phase diagram of the mono-axial chiral
helimagnet in the magnetic field plane by solving the model without any uncontrolled 
approximation, and the results which come out are in some respects rather different from those of 
reference~\onlinecite{Chapman14}. 

The paper is organized as follows. In section~\ref{sec:model} the model is described and the method 
of solution is outlined. Sec.~\ref{sec:phd} is devoted to analyze the results for the phase diagram.
In Sec.~\ref{sec:scaling} the nature of the singularities that appear on the transition line
is studied. The effect of easy-plane anisotropy is discussed in Sec.~\ref{sec:anisot}.
A comparison with approximate solutions, namely the conical approximation of 
Ref.~\onlinecite{Chapman14} and the decoupling approximation developed in this work, is presented in 
Sec.~\ref{sec:approx}, where it is also shown that the form of the phase transition line is universal
when the magnitude of the anisotropy is large. Sec.~\ref{sec:exp} deals with the application of the
results to \CrNbS. And some conclusions are gathered in Sec.~\ref{sec:conc}.
Finally, details on the numerical methods are given in the appendix.

\section{Model and method of solution \label{sec:model}}
 
To settle the question raised in the introduction, we studied the zero temperature ground state of a 
classical spin system with FM isotropic exchange, mono-axial DM interactions, and single-ion 
easy-plane anisotropy, in the presence of an external magnetic field.

At zero temperature the spins lying on the planes perpendicular to the DM axis are FM arranged.
The competition between the DM and FM interaction gives rise to a helicoidal spin structure
along the DM axis and the model becomes effectively one dimensional. The classical effective 
1D hamiltonian in the continuum limit reads:
\begin{equation}
\mathcal{H} =  \int_0^\Lambda dz \left[\frac{1}{2}\hat{n}^{\prime 2}
-q_0\hat{z}\cdot\left(\hat{n}\times \hat{n}^\prime\right) 
+ \gamma n_z^2 - \vec{\beta}\cdot\hat{n} \right] \label{hamil}
\end{equation}
where $z$ labels the coordinate along the DM axis $\hat{z}$, the prime stands for the derivative
with respect to $z$, $\hat{n}$ is a unit vector in the direction of the spin $\vec{S}$, so that 
$\vec{S}=S\hat{n}$, with $S$ being the spin modulus, and $\Lambda$ is the system length along the
DM axis, which is assumed to be large. The constants entering Eq.~(\ref{hamil})
are related to the interaction strengths as follows: let $J$ and $D$ stand for the strengths of
the FM and DM interactions, respectively, $K$ for the strength of the single-ion anisotropy 
interaction, and $a$ for the underlying lattice parameter; the energy $\mathcal{H}$ is measured 
in units of $JS^2/a$, and therefore
$q_0=D/Ja$, $\gamma=K/Ja^2$, and $\vec{\beta}=(g\mu_\mathrm{B}/JSa^2)\vec{H}$, where $\vec{H}$
is the applied magnetic field. The parallel magnetic field is thus proportional to $\beta_z$ and, 
without any loss, we can take the perpendicular field proportional to $\beta_x$ and set 
$\beta_y=0$. The parameter $q_0$ is the helix pitch at zero magnetic field, so that
its period is $L_0=2\pi/q_0$. 
We set $q_0=1$ in the computations, what merely amounts to set the unit length equal 
to $L_0/2\pi$.

\begin{figure}[t!]
\centering
\begin{minipage}{0.25\textwidth}
\centering
\includegraphics[width=0.92\linewidth,angle=0]{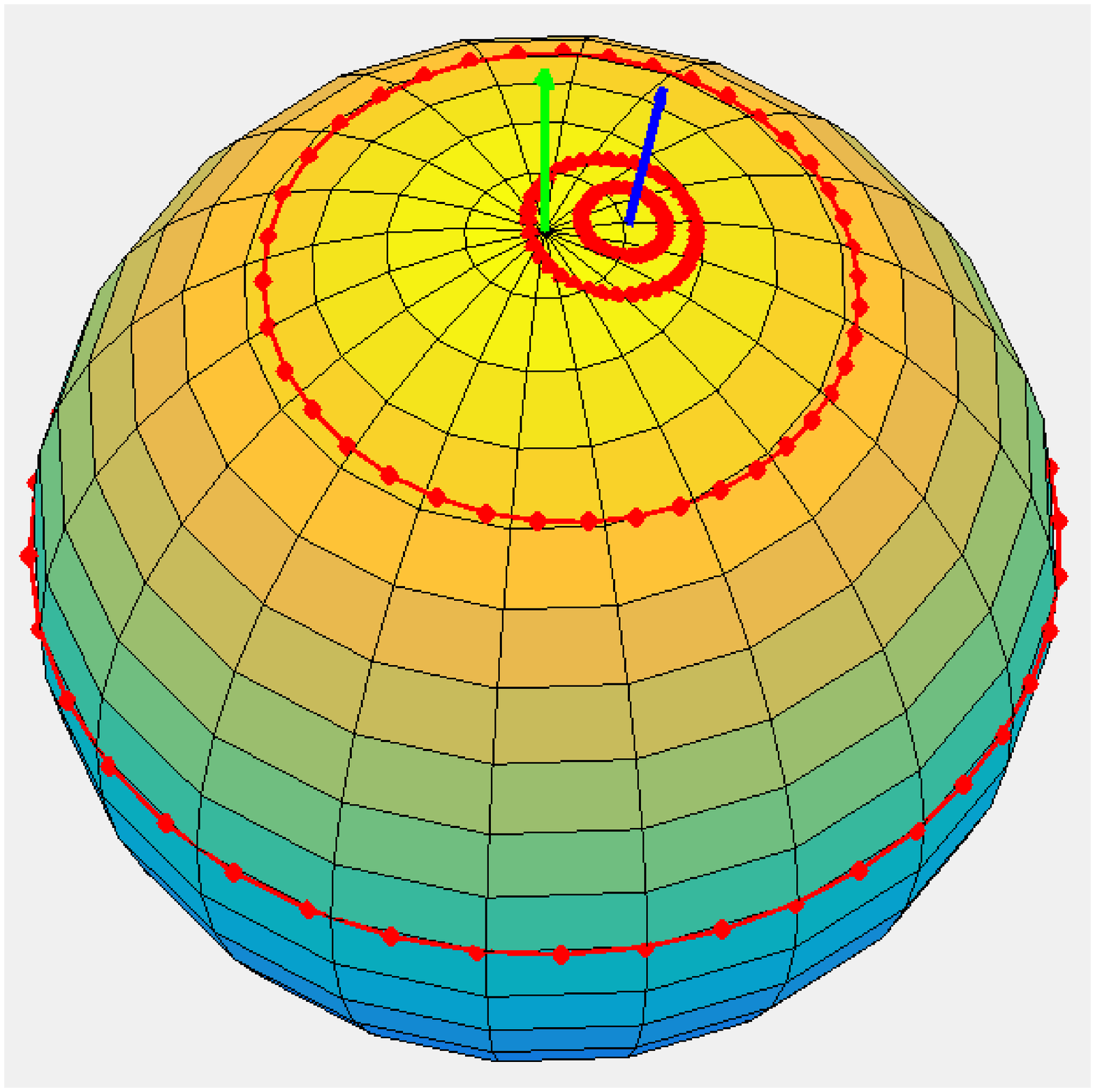}
\end{minipage}%
\begin{minipage}{0.25\textwidth}
\centering
\includegraphics[width=0.9\linewidth,angle=0]{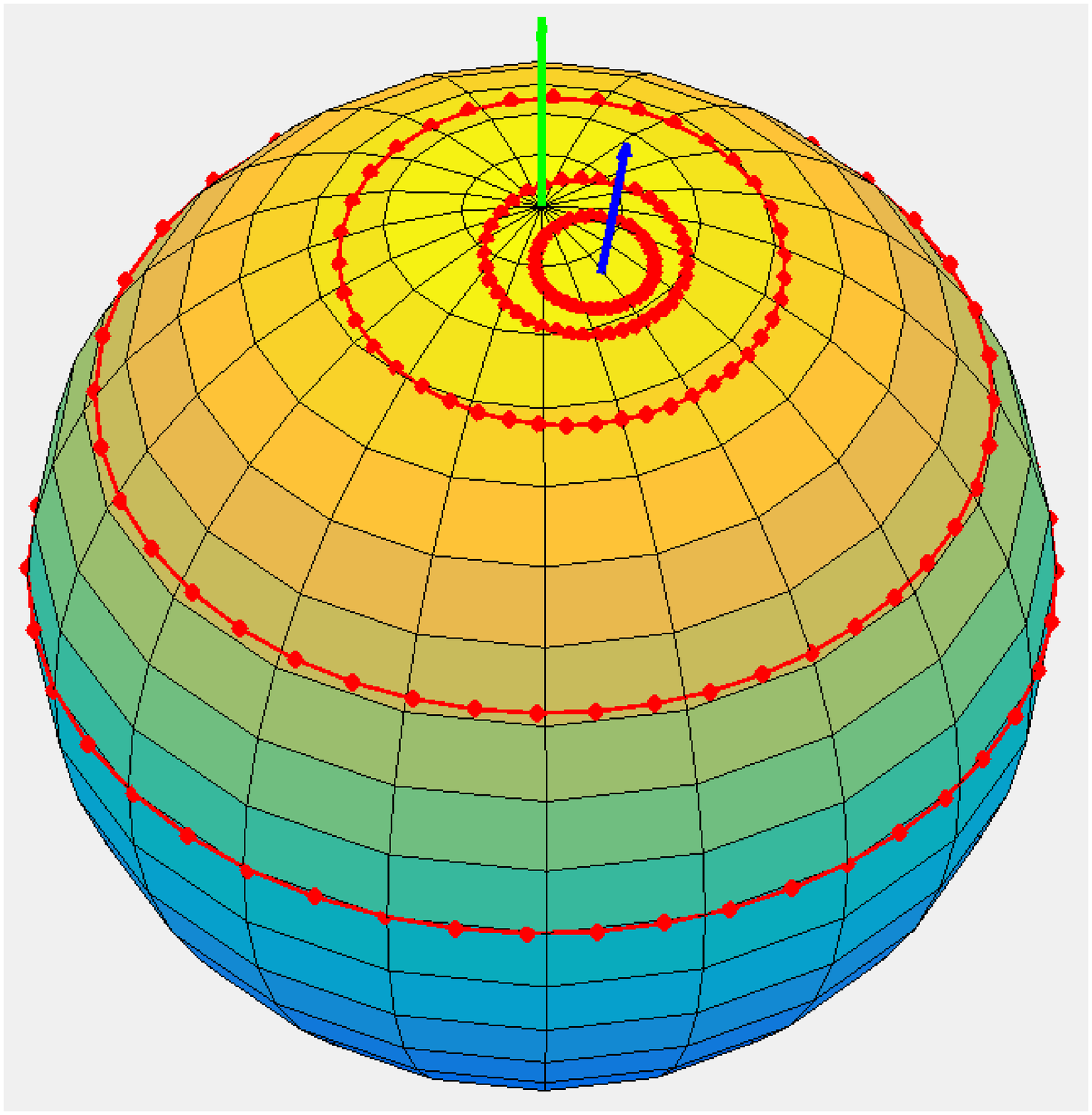}
\end{minipage}
\begin{minipage}{0.25\textwidth}
\centering
\includegraphics[width=0.92\linewidth,angle=0]{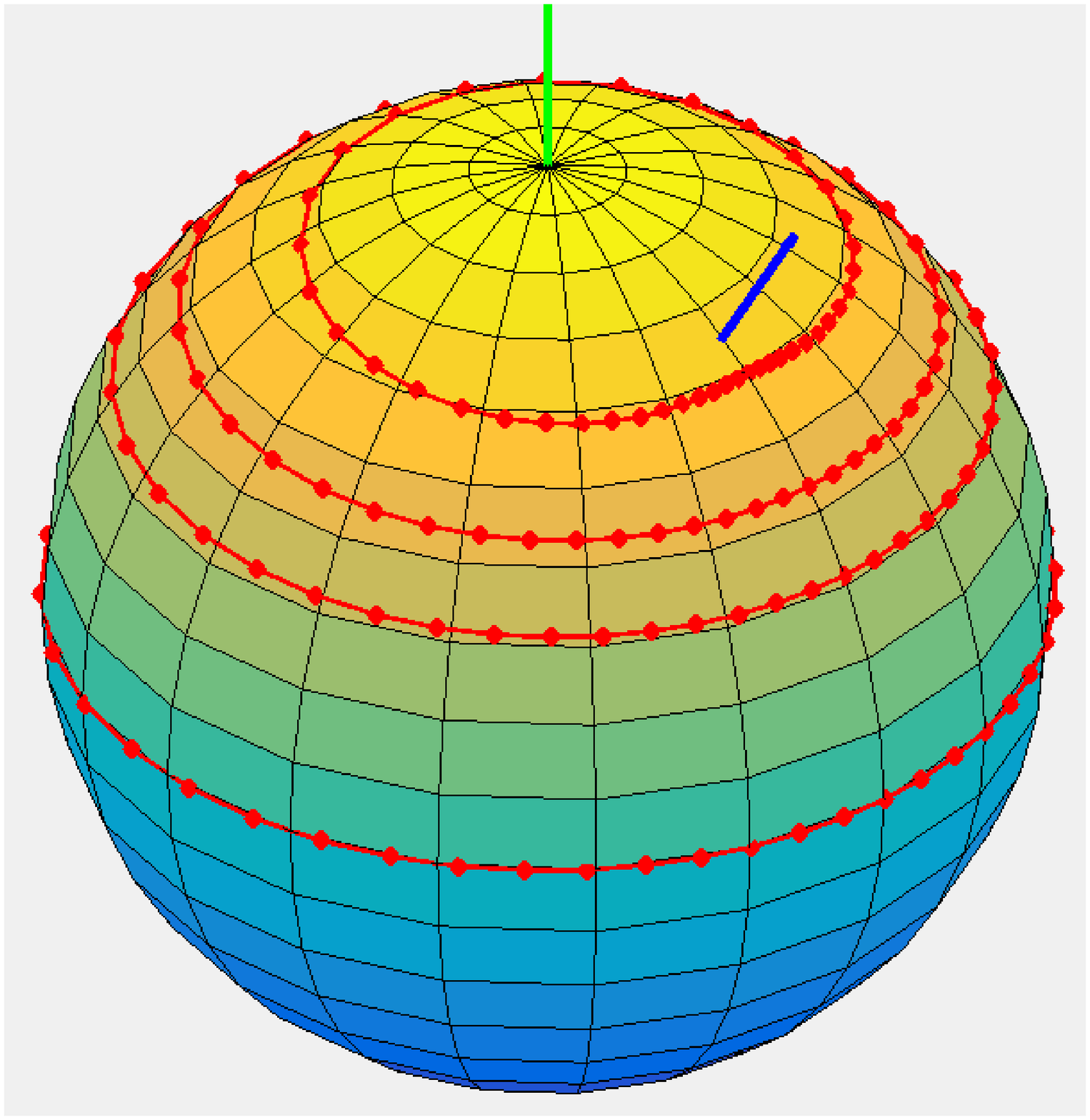}
\end{minipage}%
\begin{minipage}{0.25\textwidth}
\centering
\includegraphics[width=0.9\linewidth,angle=0]{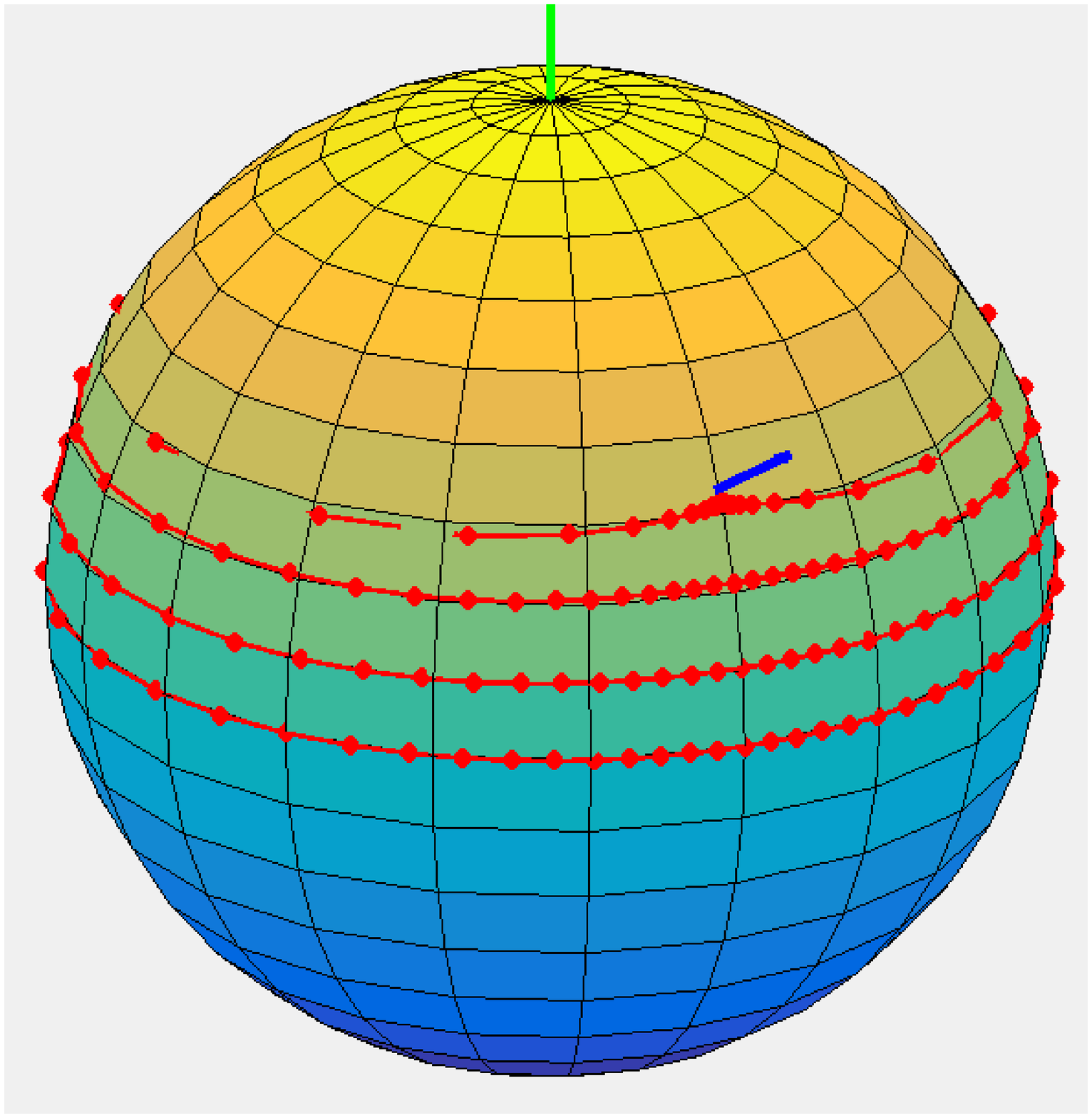}
\end{minipage}
\caption{The paths defined by the ground state configurations on the unit sphera (red curves)
in different typical cases. Top left: $\beta_x=0.15$, with transition point on CTL1.
Top right: $\beta_x=0.17$, with transition point on DTL.
Bottom left: $\beta_x=0.35$, with transition point on DTL.
Bottom right: $\beta_x=0.55$, with the transition point on CTL2. 
In each case the different curves correspond to different values of $\beta_z$.
The green and blue axis represent respectively the DM axis and the direction of the FFM state 
at the transition point.
\label{fig:sphere}}
\end{figure}

At zero temperature the classical ground state is given by the minimum of the 
hamiltonian~(\ref{hamil}), which is a solution of the corresponding Euler-Lagrange (EL) equations. 
The infinitely many solutions are characterized by the boundary conditions (BC). On physical grounds, 
we expect a periodic ground state for an infinite system. Then, if $L$ is the period, the appropriate 
BC is $\hat{n}(0)=\hat{n}(L)$. This however does not guarantee periodicity because, as the EL equations 
are second order, periodicity requires also the equality of the first derivatives,
\begin{equation}
\hat{n}^\prime(0)=\hat{n}^\prime(L)\,, \label{eq:per} 
\end{equation}
a condition which cannot be generally imposed since the 
associated boundary value problem (BVP) would be overdetermined. Hence, the value of the spin at the 
boundaries, $\hat{n}(0)$, has to be tuned in order to obtain periodicity with period $L$. 
This can in principle be done as we have the two degrees of freedom associated to $\hat{n}(0)$ and 
only two of the three Eqs.~(\ref{eq:per}) are independent, due to the constraint 
$\hat{n}(z)\cdot\hat{n}^\prime(z)=0$. The periodic solutions are thus solely characterized by the period $L$.
As the solution of a BVP is not necessarily unique, there might be several periodic solutions associated 
to a given $L$. We do not expect this on physical grounds, and indeed we never found more than one 
solution for the BVP in the course of our numerical computations. 
The equilibrium period is obtained by minimization of the energy density, 
$\mathcal{E}=\mathcal{H}/\Lambda$, which for a periodic state is given by
\begin{equation}
\mathcal{E}=(1/L)\int_0^Lh(z)dz\,,
\end{equation} 
where $h(z)$ is the integrand entering Eq.~(\ref{hamil}).

The approach proposed here is a generalisation of the well known procedure of expanding the 
magnetisation in harmonic modes and selecting the mode which minimises the free energy. 
The present problem, however, is highly non-linear and all modes contribute to the solution, which
has to be found by numerical techniques.
Dzyaloshinskii \cite{Dzyal58,Dzyal64} solved analytically the case of a purely perpendicular field
using a similar approach: he got the general solution of the differential equations and 
minimized the free energy in terms of the non-trivial integration constant, the jacobian elliptic 
modulus, which is directly related to the period of the IC structure.

The form of the EL equations depends on the coordinate system chosen to parametrize $\hat{n}$.
As the whole sphere cannot be smoothly parametrized with a single coordinate set,
we used two different coordinate sets:
a) coordinate set I, $(\xi,\varphi)$, with $\xi\in (-\infty,\infty)$, and $\varphi\in [0,2\pi]$,
so that $\hat{n}=(\cos\varphi,\sin\varphi,\xi)/\rho$, where $\rho=\sqrt{1+\xi^2}$; 
and b) coordinate set II, $(\vartheta,\phi)$, with $\vartheta\in [0,\pi]$ and $\phi\in [0,2\pi]$,
so that $\hat{n}=(\sin\vartheta\sin\phi,\cos\vartheta,\sin\vartheta\cos\phi)$.
Coordinate set I is closely related to the polar coordinate set with polar axis along the DM axis 
($\hat{z}$) and the polar angle $\theta$ related to $\xi$ by $\cot\theta=\xi$. 
This coordinate set is singular on $\pm\hat{z}$, where $\xi$ diverges. 
Coordinate set II is a polar coordinate system with polar axis along $\hat{y}$, and therefore is 
singular at $\pm\hat{y}$, where $\sin\vartheta=0$. The EL equations and the BC
read in coordinate set I

\begin{widetext}
\begin{subequations}
\label{eq:EL1}
\begin{eqnarray}
&&\xi^{\prime\prime} - 2\xi\xi^{\prime2}/\rho^2 + (\varphi^{\prime2}-2q_0\varphi^\prime)\xi - 2\gamma\xi
- \beta_x\rho\xi\cos\varphi + \beta_z\rho = 0\,, \label{eq:EL1a} \\ 
&&\varphi^{\prime\prime} - 2(\varphi^\prime-q_0)\xi\xi^\prime/\rho^2 
- \beta_x\rho\sin\varphi = 0\,, \label{eq:EL1b} \\
&&\varphi(0)=0, \;\;\;\;\; \varphi(L)=2\pi, \;\;\;\;\; \xi(0)=\xi(L)=\xi_0\,;
\label{eq:BC1} 
\end{eqnarray}
\end{subequations}
and in coordinate set II
\begin{subequations}
\label{eq:EL2}
\begin{eqnarray}
&&\vartheta^{\prime\prime} - \sin\vartheta\cos\vartheta\,\phi^{\prime2}
+2q_0\sin^2\vartheta\cos\phi\,\phi^{\prime}
-2\gamma\sin\vartheta\cos\vartheta\cos^2\phi
+\beta_x\cos\vartheta\sin\phi + \beta_z\cos\vartheta\cos\phi = 0\,,
\label{eq:EL2a} \\ 
&&\sin^2\vartheta\,\phi^{\prime\prime} + \sin\vartheta\cos\vartheta\,\vartheta^\prime\phi^\prime
- 2q_0\sin^2\vartheta\cos\phi\,\vartheta^\prime
+2\gamma\sin^2\vartheta\sin\phi\cos\phi 
+ \beta_x\sin\vartheta\cos\phi - \beta_z\sin\vartheta\sin\phi = 0\,, \label{eq:EL12b} \\
&&\vartheta(0)=\vartheta(L)=\pi/2\,, \;\;\;\;\; \phi(0)=\phi(L)=\phi_0\,. \label{eq:BC2}
\end{eqnarray}
\end{subequations}
\end{widetext}

As $\hat{n}$ contains two degrees of freedom and the differential equations are second order, 
the general solution contains four independent constants. The condition $\hat{n}(0)=\hat{n}(L)$ 
removes two of them. We used translational invariance to eliminate one more, so that it remains 
either $\xi_0$ or $\phi_0$ as tuneable parameter. The reason is that the configuration 
that minimizes the energy will have at least one point with the spin lying on the plane determined 
by the DM axis ($\hat{z}$) and the magnetic field ($\hat{x}$). Using translational symmetry we 
can choose this point as $z=0$, so that in coordinate set I we have $\varphi(0)=0$ and in 
coordinate set II $\vartheta(0)=\pi/2$.

The ground state (or equilibrium) configuration $\hat{n}(z)$ with $z\in[0,L]$ can be visualized as a
closed path on the unit sphere, with the coordinate $z$ acting as the parameter of the curve.
Fig.~\ref{fig:sphere} displays such paths, given by the red trajectories, in different situations. 
The green axis represents the DM axis ($\hat{z}$) and the FFM
state at the transition point is represented by the blue axis. Coordinate set I and its 
associated BC is appropriate when the path does encircle the DM axis; 
coordinate set II is suitable when the 
paths are not too close to $\hat{y}$, which lies on the equatorial plane. 
As already mentioned, notice that the BC~(\ref{eq:BC1}) forces the path to encircle the DM axis. 
When the values of the magnetic field are such that the equilibrium state is represented
by one path which does not encircle the DM axis, as in the top panels of
Fig.~\ref{fig:sphere}, the BVP~(\ref{eq:EL1}) has no solution, and we are forced to use
coordinate set II and solve~(\ref{eq:EL2}). On the other hand, on a wide part of the phase diagram
both coordinate sets can be used. In such cases, we solved both BVPs and got the same solution
$\hat{n}(z)$ within the tiny numerical uncertainties (see the appendix, where details on the 
numerical procedures are given).

\section{Phase diagram \label{sec:phd}}

To obtain a phase diagram we compare the energies of the IC state, $\mathcal{E}_\mathrm{IC}$, which is
computed numerically following the lines outlined in the previous section, and of the FFM state,
$\mathcal{E}_\mathrm{FFM}$, for which $\varphi=0$ and the constant value of $\xi$ is given by the 
solution of
\begin{equation}
\frac{\xi}{\sqrt{1+\xi^2}} = \frac{\beta_z}{2\gamma+\beta_x\sqrt{1+\xi^2}}\,.
\end{equation}

For small $\vec\beta$, 
$\mathcal{E}_\mathrm{FFM}>\mathcal{E}_\mathrm{IC}$ and the IC state is the ground state. The transition 
to the FFM state takes place when $\mathcal{E}_\mathrm{FFM}=\mathcal{E}_\mathrm{IC}$, and is continuous 
if at this point the IC state merges smoothly with the FFM. Otherwise the transition is discontinuous. 
The IC state can be visualized as a closed curve on the unit sphere while the FFM state is represented 
by a single point (Fig.~\ref{fig:sphere}).
As discussed in the introduction, there are two mechanisms by which the IC state can be continuously
transformed into the FFM state. The first possibility is displayed in the top left panel of 
Fig.~\ref{fig:sphere}: as the magnetic field is tuned to its critical value the IC curve 
reduces its size until it collapses onto the FFM state. In this case a helical conical state which,
in the vicinity of the transition point, revolves around the direction of the FFM and not around the
DM axis, becomes completely parallel at the transition point. The bttom right panel of 
Fig.~\ref{fig:sphere} illustrates the second possibility. The length of the IC curve on the unit sphere
remains finite as the transition point is approached. Near the transition, the vast majority of the spins, 
however, are concentrated on a narrow arc close to the FFM state and the number of spins lying on the 
remaining part of the curve becomes negligible, so that a CSL is formed. At the transition point the
period of the soliton diverges and therefore its fundamental wavevector, $q=2\pi/L$, tends to zero. 

\begin{figure}[t!]
\centering
\includegraphics[width=0.45\textwidth,angle=0]{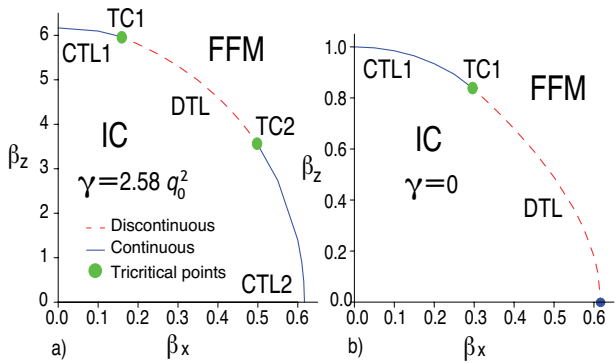}
\caption{The phase diagram in the $(\beta_x,\beta_z)$ plane with easy-plane anisotropy
$\gamma=2.584q_0^2$ a) and $\gamma=0$ b). Red dashed and blue continuous lines represent, 
respectively, discontinuous and continuous phase transition. Green points are the tricritical 
points.
\label{fig:phd}}
\end{figure}

\begin{figure}[t!]
\centering
\includegraphics[width=\linewidth,angle=0]{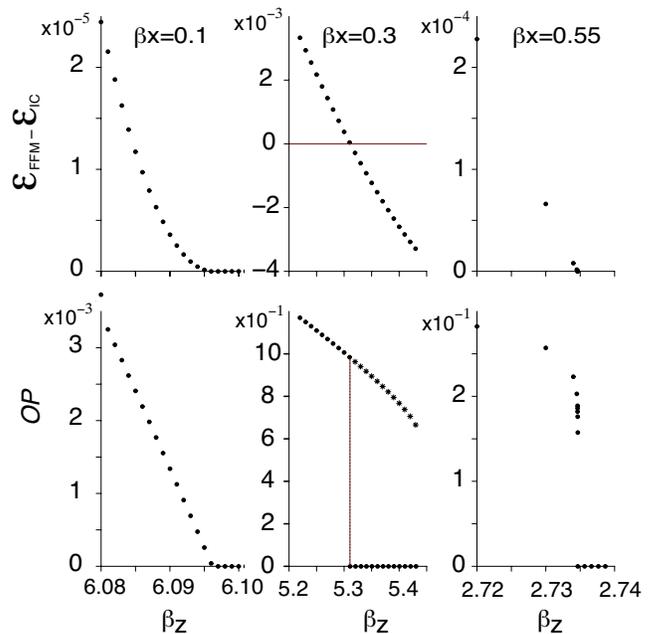}%
\caption{$\Delta\mathcal{E}=\mathcal{E}_\mathrm{FFM}-\mathcal{E}_\mathrm{IC}$ and $\mathcal{OP}=1-M$ 
for three values of $\beta_x$, for a system with easy-plane anisotropy $\gamma=2.584q_0^2$. 
\label{fig:trans}}%
\end{figure}

Fig.~\ref{fig:phd} displays the phase diagram in the $(\beta_x,\beta_z)$ plane for
$\gamma=2.584 q_0^2$ (a) and $\gamma=0$ (b). We shall disccuss the $\gamma=0$ case in 
section~\ref{sec:anisot}
and here the $\gamma=2.584 q_0^2$ case, which is relevant for \CrNbS\ (see section~\ref{sec:exp}). 
We see that the transition is continuous in the vicinities of the $\beta_x=0$ and the 
$\beta_z=0$ axes, while it is discontinuous in the intermediate regime.  
The behaviour of $\mathcal{E}_\mathrm{FFM}-\mathcal{E}_\mathrm{IC}$
as a function of $\beta_z$ for fixed $\beta_x$
is illustrated in the top panels of Fig.~\ref{fig:trans} in three cases, one for each of the three 
transition lines. 
We used $\mathcal{OP}=1-M$, where $M=|(1/L)\int_0^L\hat{n}dz|$ is the magnetization in suitable
units, as an ``order parameter'' (OP) which vanishes in the FFM phase.
Its behaviour in each case is displayed in the bottom panels of Fig.~\ref{fig:trans}.

For conciseness, we shall call the continuous transition lines
which touch the $\beta_x=0$ and the $\beta_z=0$ axes respectively the continuos transition line 
1 (CTL1) and 2 (CTL2), and we shall use DTL for discontinuous transition line.
On the CTL1 the FFM state is reached continuously by the ``closing cone'' mechanism, as illustrated
in Fig.~\ref{fig:sphere} (top left). The period $L$ remains finite and close to $L_0$
at the transition point.
On the other hand, on the CTL2, the period of the IC diverges and the FFM state is 
continuously reached by the same mechanism as in the $\beta_z=0$ case. The bottom right panel of
 Fig.~\ref{fig:sphere} is an instance of this mechanism.

\begin{figure}[t!]
\centering
\begin{minipage}{0.25\textwidth}
\centering
\includegraphics[width=\linewidth,height=\linewidth,angle=0]{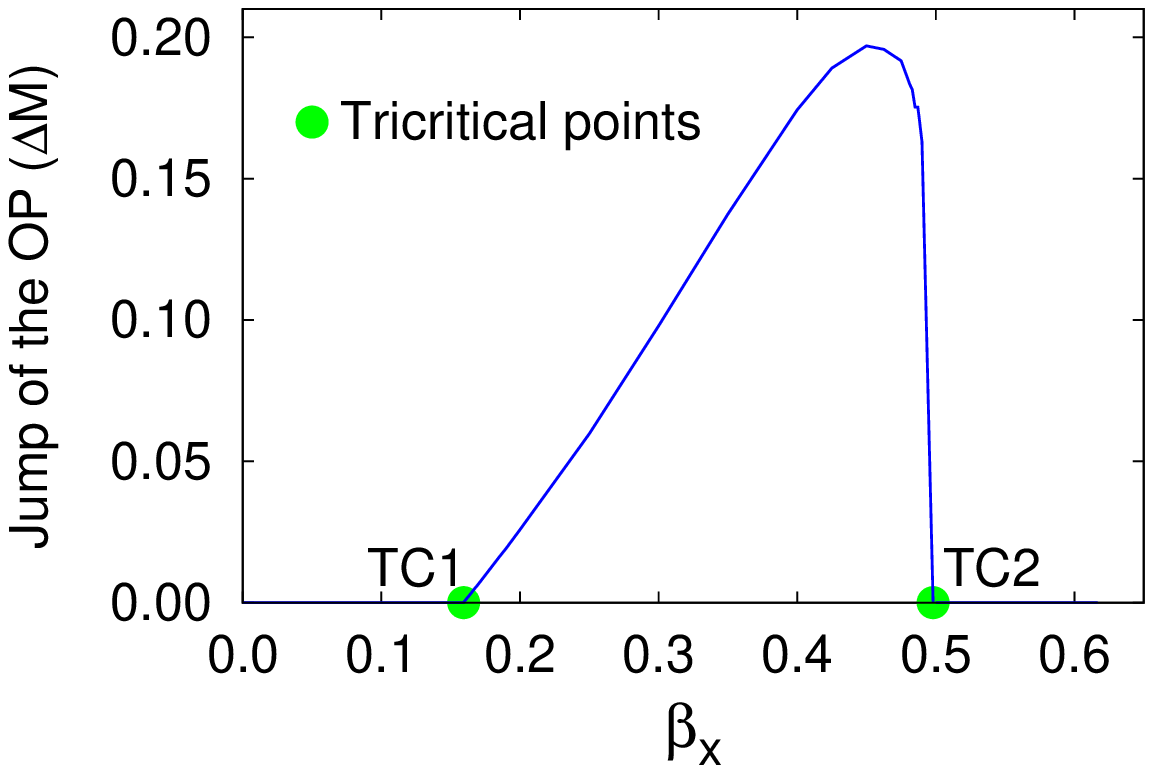}%
\end{minipage}%
\begin{minipage}{0.25\textwidth}
\centering
\includegraphics[width=\linewidth,height=0.9\linewidth,angle=0]{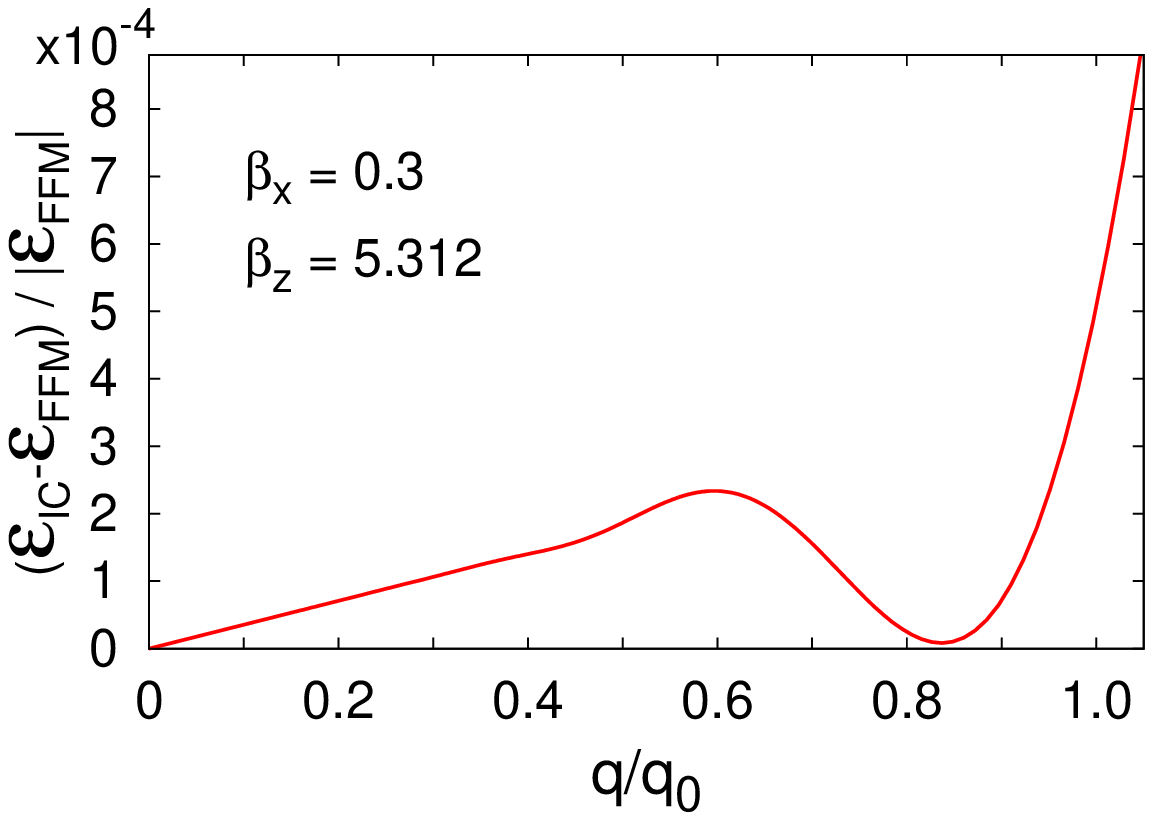}%
\end{minipage}
\caption{Left: the order parameter jump along the transition line parametrised by 
 $\beta_x$. Right: the excess of energy of the IC state over the FFM state as a function of the 
fundamental wavevector $q=2\pi/L$; in this 
case the FFM state ($q=0$) is stable and the IC state with $q\approx 0.84\,q_0$ is metastable. 
In both figures the easy-plane anisotropy is $\gamma=2.584\,q_0^2$.
\label{fig:OP}}%
\end{figure}

On the DTL the IC and FFM state become degenerate and neither the ``cone closes'' 
nor the period diverges. These situations are illustrated in the top right and bottom left panels of 
Fig.~\ref{fig:sphere}. The OP is discontinuos on this line. Fig.~\ref{fig:OP} (left) shows the 
OP jump along the transition line, parametrised by $\beta_x$.
Along the DTL the IC state coexists with the FFM state as they are degenerate in energy.
On each side of the line either the IC or the FFM state are metastable. This is illustrated
in Fig.~\ref{fig:OP} (right), where the energy difference is displayed as a function of $q=2\pi/L$
in a case in which the IC state with $q\approx 0.84\,q_0$ is metastable and the FFM ($q=0$) is stable.
Hence, phenomena like phase coexistence, domain formation, and hysteresis are expected
in some regions of the magnetic phase diagram. The metastability does not cause 
numerical problems since $L$ is fixed in the numerical computations, and the BC~(\ref{eq:BC1})
prevent the switching between the IC and the FFM states.

The DTL is separated from the CTL1 and the CTL2 by two tricritical points, called 
respectively TC1 and TC2. In Fig.~\ref{fig:OP} (left) these are the points where the OP jump 
ceases to vanish, and are signaled by the two green circles. 
For $\gamma=2.584q_0^2$ TC1 is located at $(\beta_{x}^{t1},\beta_{z}^{t1})=(0.1593,5.9552)$ 
and TC2 at $(\beta_{x}^{t2},\beta_{z}^{t2})=(0.498,3.5301)$. 

\begin{figure}[t!]
\centering
\includegraphics[width=0.7\linewidth,height=0.52\linewidth,angle=0]{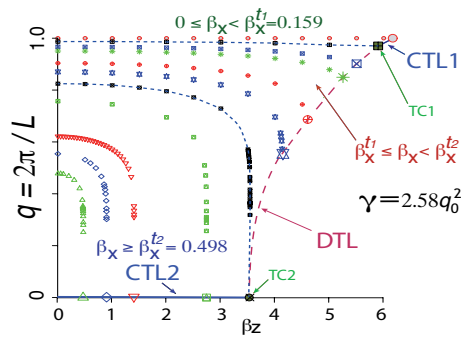}%
\caption{The inverse of the period of the IC state as a function of $\beta_z$  for several 
values of $\beta_x$. The critical $\beta_z$ for each $\beta_x$ is given by the point with 
minimum $q$ in the corresponding curve. 
The two tricritical points are given by the points with minimum $q$ on each of the two
dashed lines ($\beta_x^{t1}=0.159$ and $\beta_x^{t2}=0.498$) with black symbols. 
\label{fig:L}}
\end{figure}

Fig.~\ref{fig:L} displays the fundamental wavevector $q$ as a function of $\beta_z$ for different 
values of $\beta_x$, including $\beta_x^{t1}=0.159$ and $\beta_x^{t2}=0.498$ 
(dashed lines with black symbols). Notice that $q$ is almost constant ($q\approx q_0$) for
$\beta_x<\beta_x^{t1}$, while it decreases with $\beta_x$ and with $\beta_z$ for
$\beta_x>\beta_x^{t1}$. The variation of $q$ becomes very abrupt close to the tricritical
point TC2.

\begin{figure}[t!]
\centering
\begin{minipage}{0.25\textwidth}
\centering
\includegraphics[width=\linewidth,angle=0]{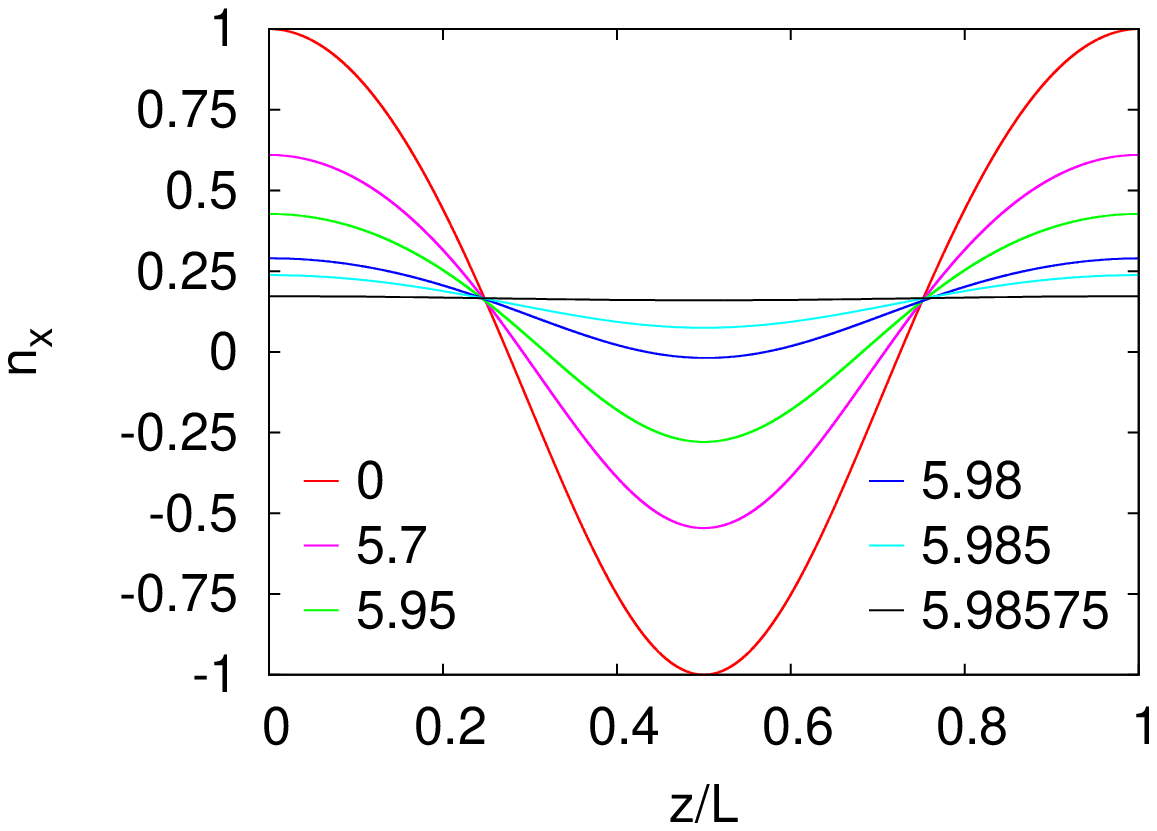}%
\end{minipage}%
\begin{minipage}{0.25\textwidth}
\centering
\includegraphics[width=\linewidth,angle=0]{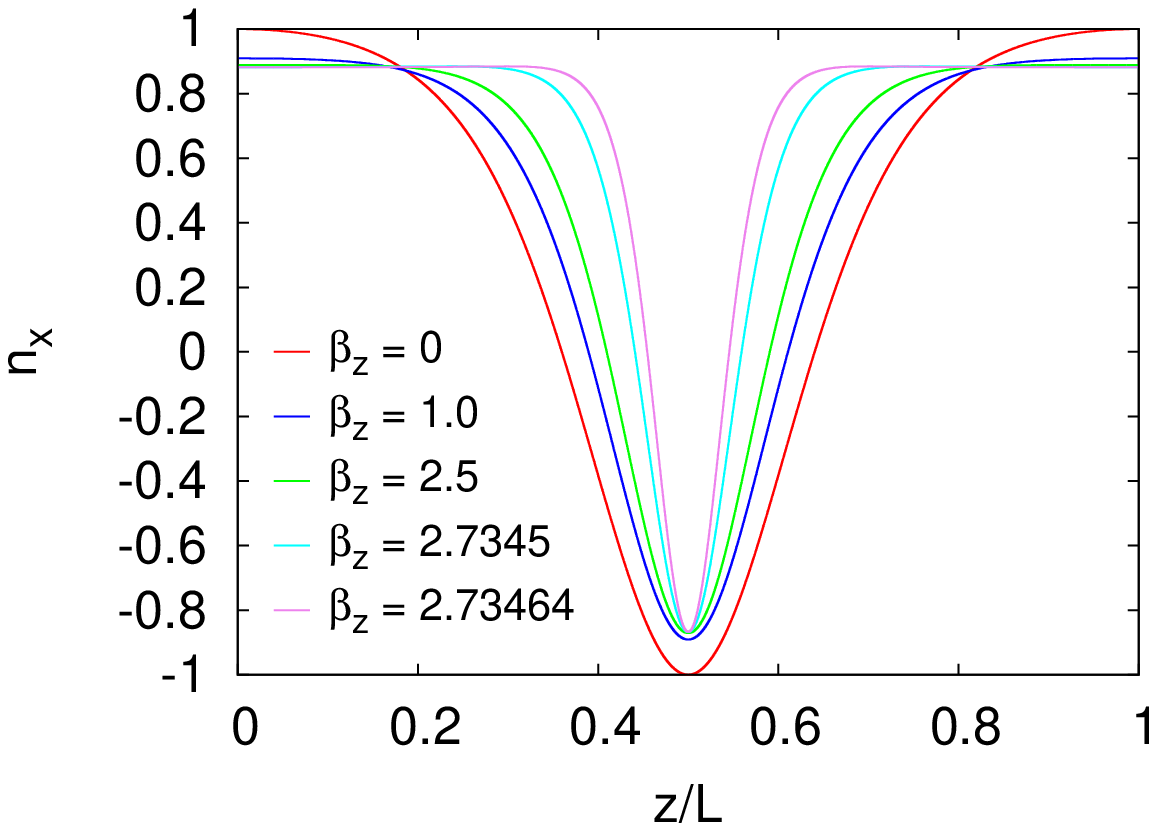}%
\end{minipage}

\begin{minipage}{0.25\textwidth}
\centering
\includegraphics[width=\linewidth,angle=0]{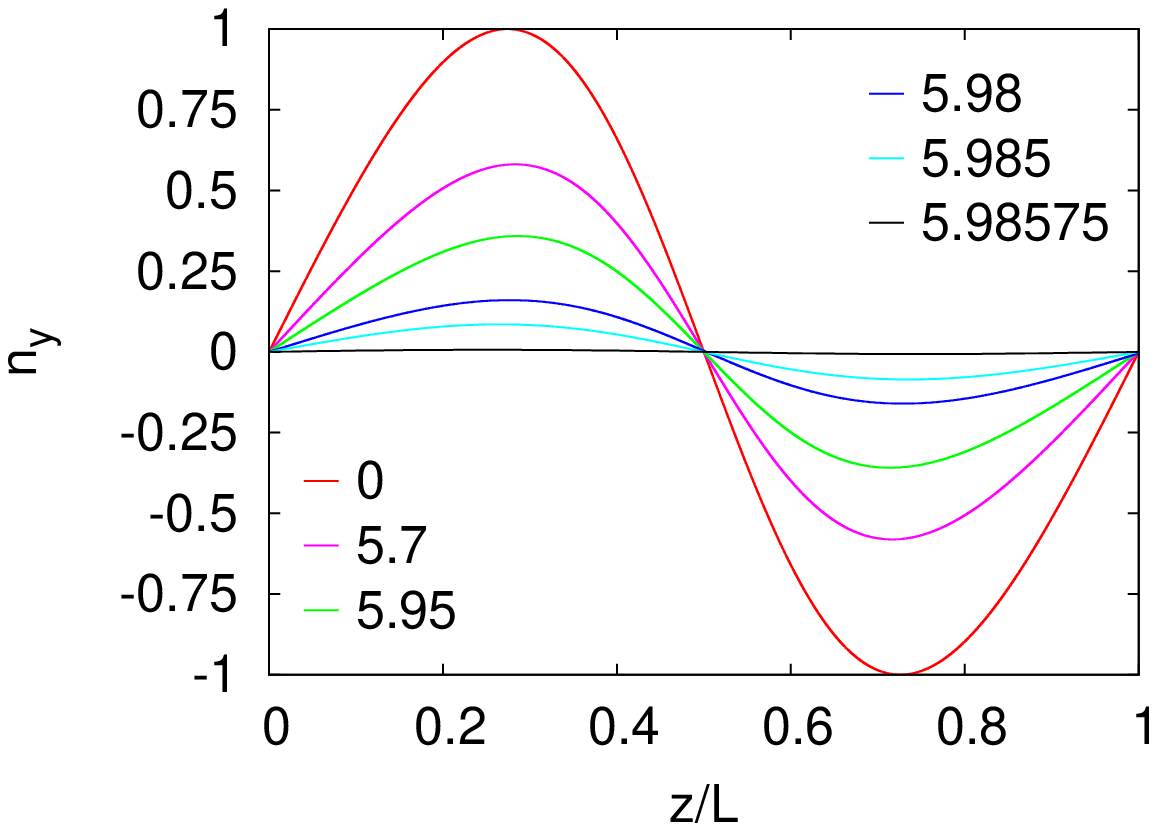}%
\end{minipage}%
\begin{minipage}{0.25\textwidth}
\centering
\includegraphics[width=\linewidth,angle=0]{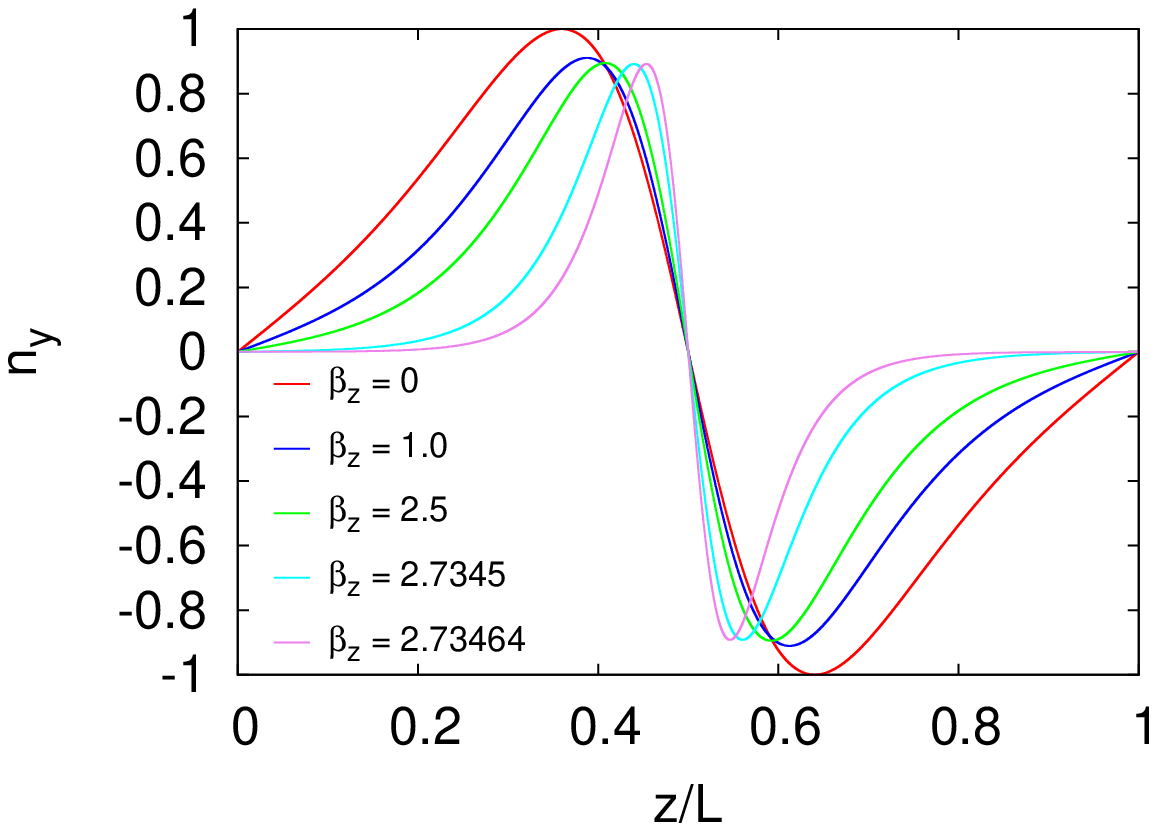}%
\end{minipage}%

\begin{minipage}{0.25\textwidth}
\centering
\includegraphics[width=\linewidth,angle=0]{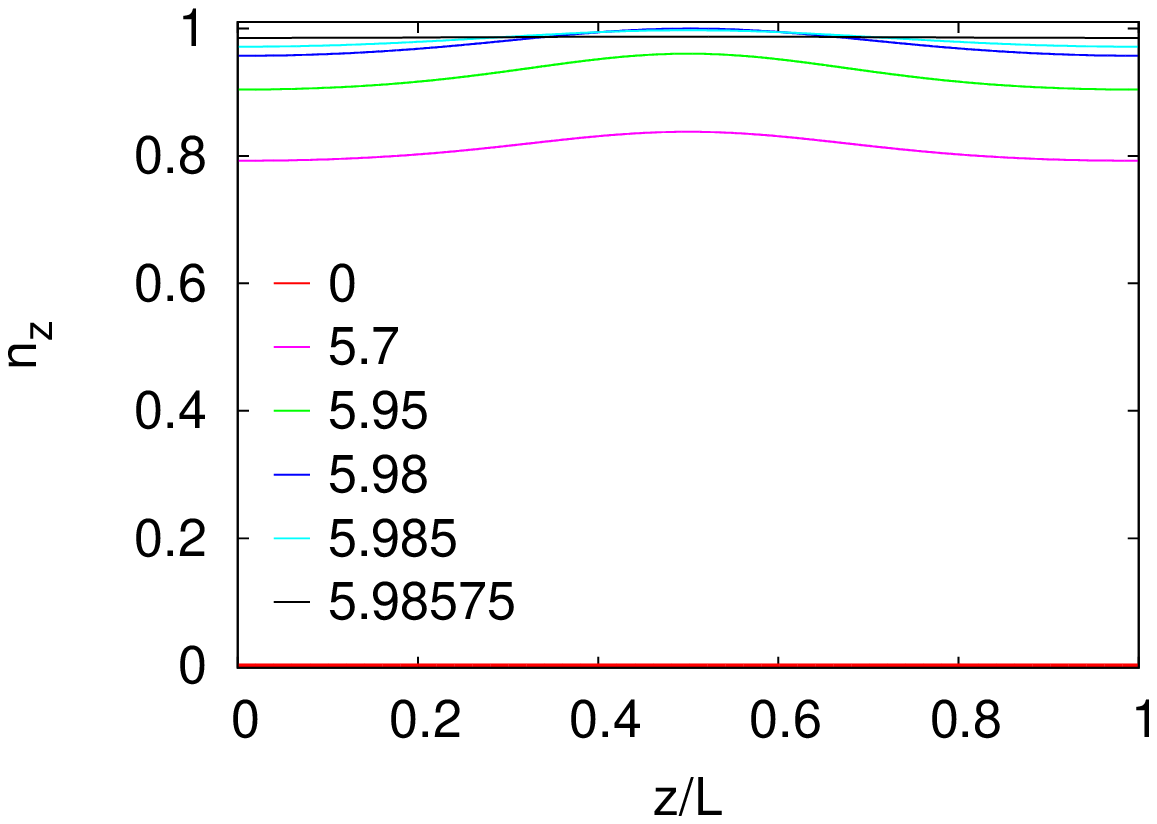}%
\end{minipage}%
\begin{minipage}{0.25\textwidth}
\centering
\includegraphics[width=\linewidth,angle=0]{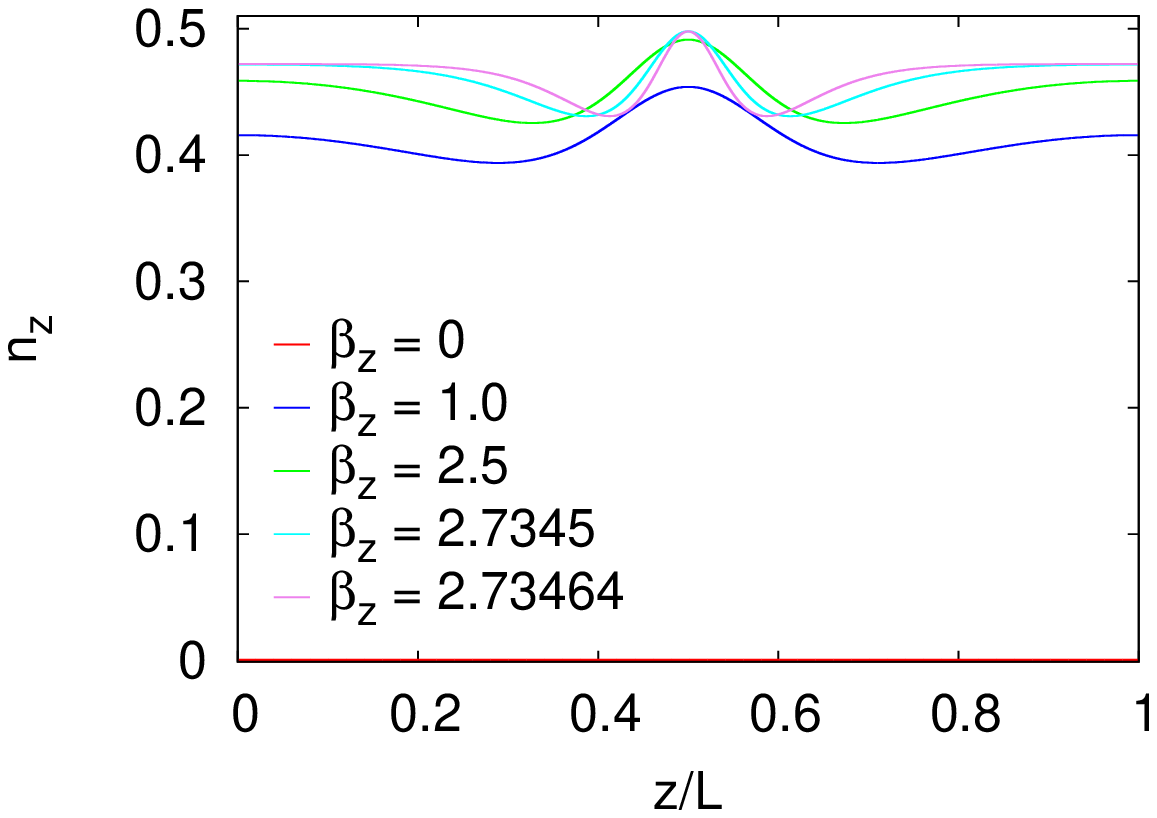}%
\end{minipage}%
\caption{
The cartesian components of $\hat{n}$ as a function of $z/L$:
$n_x$ (top), $n_y$ (middle), and $n_z$ (bottom),
for $\gamma=2.584q_0^2$ and $\beta_x=0.15$ (left) and 
$\beta_z=0.55$ (right), and different values of $\beta_z$ approaching
the transition point. The DM axis and the magnetic field point in the 
direction of $\hat{z}$ and $\hat{x}$, respectively.
\label{fig:configs}}%
\end{figure}

The spatial dependence of the ground state as the transition is approached is shown in
Fig.~\ref{fig:configs}, where the cartesian components of $\hat{n}$ are displayed as a function
of $z/L$. Cartesian components are shown since they are regular for any $\vec{\beta}$.
They have been computed from the solutions obtained either with coordinate set I or II.
The left panels correspond to $\beta_x=0.15$ and show $\hat{n}(z)$ for different values of
$\beta_z$ which tend to a transition point on CTL1. We observe that the variation of $\hat{n}(z)$
is distributed smoothly over the whole period for all $\beta_z$, so that no soliton is formed.
Also, the amplitude of the oscilations decreases smoothly to zero as the transition point is
approached, attaining continuously the FFM.
The right panels correspond $\beta_x=0.55$ and the values of $\beta_z$ tend to a transition
point on CTL2. Observe that the variations of $\hat{n}(z)$ are gradually concentrated on a narrow
section of the period, and a soliton lattice is formed. It is a relative narrowness: the size of
the region where $\hat{n}(z)$ varies noticeably is approximately constant, but the period increases
as the transition is approached.
It is interesting that the soliton lattice can be created by increasing the component of the 
magnetic field parallel to the DM axis, keeping constant the perpendicular field. 

\section{Singular behaviour on the continuous transition lines \label{sec:scaling}}

On the continuous transition lines CTL1 and CTL2 the IC and FFM states merge continuously but,
as in any phase transition, physical observables may present peculiar singularities. 
Let us analyze the nature of these singularities.
The OP vanishes linearly as the CTL1 ($\beta_x<\beta_x^{t1}$) is attained from the 
IC phase (bottom left pannel of Fig.~\ref{fig:trans}). Thus, no singularity appears on CTL1.
As TC1 is approached, the slope of the OP becomes more abrupt and becomes singular at TC1. 
We assume that at TC1 the OP vanishes as a power law,
$\mathcal{OP}\sim A(\beta_{z}^{t1}-\beta_z)^{n/m}$, with $n$ and $m$ small integers. 
The best fit, which is indeed very good, gives $\beta_{z}^{t1}\approx 5.9552506$ 
and $n/m=0.5002\approx 1/2$. 

\begin{figure}[t!]
\centering
\includegraphics[width=\linewidth,angle=0]{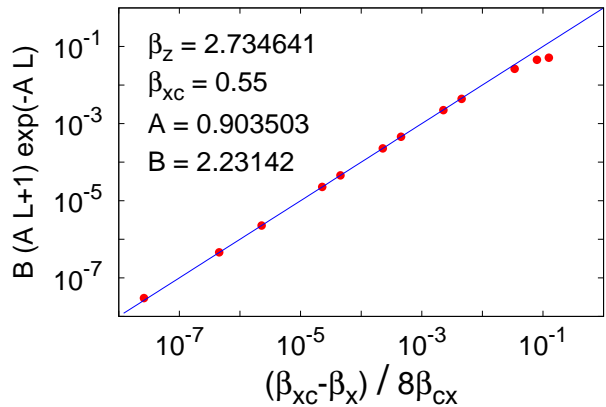}%
\caption{Scaling of the period (L) as CTL2 is approached along the 
$\beta_z=2.734641$ line for $\gamma=2.584q_0^2$. 
The blue line is $y=x$ in a $(x,y)$ plane. \label{fig:scaling}}%
\end{figure}

As CTL2 is attained from the IC phase, the OP vanishes more abruptly than a power (bottom right 
pannel of Fig.~\ref{fig:trans}). The OP scales with the period as $1/L$, so that the OP singularity
is directly related to the $L$ singularity.
For $\beta_z=0$ the following scaling law is derived from the analytically known solution:
\begin{equation} 
(\sqrt{\beta_{xc}}L+1)\exp(-\sqrt{\beta_{xc}}L)\sim (\beta_{xc}-\beta_x)/8\beta_{xc}\,, 
\end{equation}
where $\beta_{xc}=(\pi^2/16)q_0^2$ is the critical field at $\beta_z=0$. Along CTL2, 
($\beta_x^{t2}<\beta_x\leq\beta_{xc}$), a generalization of the above scaling law, 
given by 
\begin{equation} 
B(AL+1)\exp(-AL)\sim (\beta_{c}-\beta)/8\beta_{c}\,, 
\label{eq:scaling}
\end{equation}
holds. In this expression $\beta$ stands either for $\beta_x$ or for $\beta_z$, and the other 
component of $\vec{\beta}$ has to be kept constant.
The parameters $A$ and $B$ vary smoothly along CTL2. The value of $B$ depends
on whether CTL2 is attained keeping $\beta_x$ or $\beta_z$ constant, but $A$, which 
characterises the singularity, does not. Fig.~\ref{fig:scaling} displays an example of the 
scaling of $L$ as the transition point ($\beta_x=0.55$, $\beta_z=2.734641$) is approached
along the line of constant $\beta_z$. The parameter $A$ grows as $\beta_x$ decreases and
diverges at the tricritical point TC2. This means the nature of the singularity changes at
the tricritical point. Unfortunately, this new singularity cannot be studied numerically without
further insight.

The scaling of $L$ given by (\ref{eq:scaling}) is a general
feature of the formation of the CSL, and it can be used to fit the experimental results near the
transition point for oblique fields. The behaviour of the parameter $A$ then can be used to locate
the tricritical point TC2.

\section{Effect of the easy-plane single-ion aniostropy \label{sec:anisot}}

\begin{figure}[t!]
\centering
\includegraphics[width=\linewidth,angle=0]{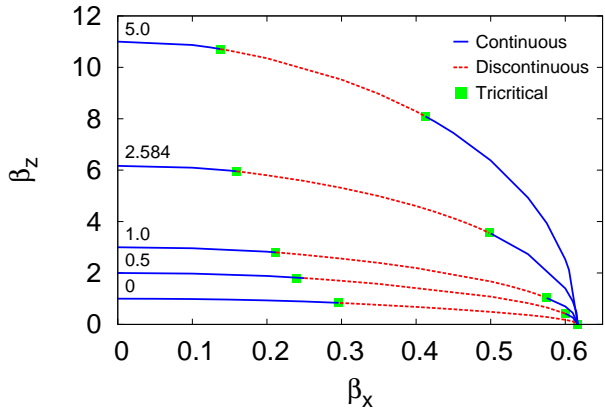}%
\caption{The transition lines in the $(\beta_x,\beta_z)$ plane for different values of the
easy-plane single-ion anisotropy $\gamma/q_0^2$. \label{fig:phd_gamma}}%
\end{figure}

The easy-plane anisotropy has an important effect on the phase diagram, as can be seen in
Fig.~\ref{fig:phd}b, where the $\gamma=0$ case is displayed. We see that the CTL2 disappears, 
the DTL reaches the $\beta_z=0$ axis, and only one tricritical point (TC1) appears.
For $\gamma>0$ the phase diagram is qualitatively similar to the case discussed in the previous
sections ($\gamma=2.584q_0^2$). The transition lines in the ($\beta_x,\beta_z$) plane for different
values of $\gamma$ are shown in Fig.~\ref{fig:phd_gamma}. By increasing $\gamma$ the lengths of 
CTL1 and CTL2 respectively decreases and increases. In a tridimensional parameter space 
$(\beta_x,\beta_z,\gamma)$ there is a phase transition surface divided  by two tricritical lines 
into three sectors, one in which the transitions are discontinuous and two in which they are 
continuous.
 
Not surprisingly, the easy-plane anisotropy provides stability to the CSL formed when the 
field is purely perpendicular, so that for large $\gamma$ large parallel fields are necessary to 
modify the behaviour of the CSL and its transition to the FFM state.
The effect of easy-plane anisotropy is important since low-anisotropy compounds described by the 
model studied in this work, as for instance thin films of MnSi~\cite{Wilson13}, will be characterized 
by low values of $\gamma$.

The case of easy-axis aniostropy ($\gamma<0$), not analyzed in this work, is theoretically 
interesting --and perhaps also phenomenologically-- as peculiar phenomena may take place in 
the crossover to Ising-like behaviour as $|\gamma|$ increases \cite{Liu73}.

\section{Comparison with simplifying approximations \label{sec:approx}}

It is interesting to compare the results obtained here with those obtained by making 
approximations which drastically simplify the mathematical problem. We consider two of such 
approximations: the conical approximation of Ref.~\onlinecite{Chapman14} and the decoupling
approximation developed below. 

In the conical approximation
the spatial variation of the spin component parallel to the DM axis, $n_z$, is neglected,
and the ground state is obtained by minimizing the energy density over spin configurations
with constant $n_z=\cos\theta_0$. The EL equations for the non-constant components
give a sine-Gordon equation similar to that of the $\beta_z=0$ model with an effective perpendicular 
field $\beta_x^{\mathrm{eff}}=\beta_x/\sin\theta_0$. Then, the ground state has the form\cite{Chapman14}
\begin{eqnarray}
n_x(z) &=& \sin\theta_0\cos\varphi(z), \label{eq:nx_cone}\\
n_y(z) &=& \sin\theta_0\sin\varphi(z), \label{eq:ny_cone}\\
n_z(z) &=& \cos\theta_0, \label{eq:nz_cone}
\end{eqnarray}
where
\begin{equation}
\varphi(z)=-2\,\mathrm{am}\left(\sqrt{\beta_x/(\kappa^2\sin\theta_0)}z,\kappa\right),
\label{eq:sg}
\end{equation}
and $\mathrm{am}(x,k)$ is the Jacobi amplitude function with elliptic modulus $\kappa$. 
The energy density associated to the above solution depends on the two parameters $\cos\theta_0$ 
and $\kappa$, and reads\cite{Chapman14}
\begin{widetext}
\begin{equation}
\mathcal{E} = \gamma\cos^2\theta_0-\beta_z\cos\theta_0 - \frac{q_0^2}{2}\sin^2\theta_0
-\beta_x\sin\theta_0\left(\frac{2}{\kappa^2}
+\frac{\pi \sqrt{q_0^2\sin\theta_0/\beta_x}}{\kappa K(\kappa)}
-\frac{4E(\kappa)}{\kappa^2K(\kappa)}-1 - \frac{q_0^2}{2\beta_x}\sin\theta_0\right),
\label{eq:ener_conical}
\end{equation}
\end{widetext}
where $K(\kappa)$ and $E(\kappa)$ are the complete elliptic integrals of the first and second
kind, respectively. The first three terms give the energy of the conical helix with wavevector 
$q_0$, while the term proportional to $\beta_x$ is the difference between the conical helicoid 
(\ref{eq:sg}) and the conical helix. We write the energy in this way for later
convenience.

\begin{figure}[t!]
\centering
\includegraphics[width=\linewidth,angle=0]{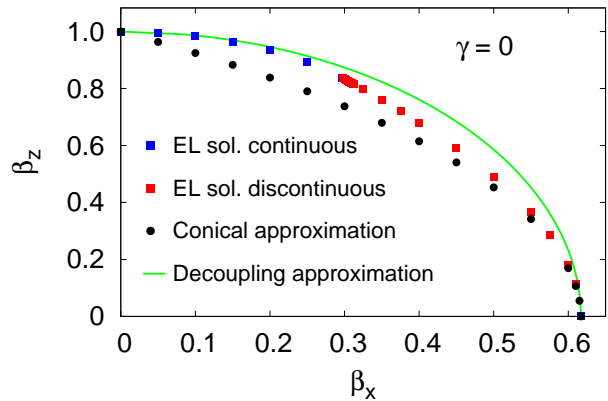}%

\includegraphics[width=\linewidth,angle=0]{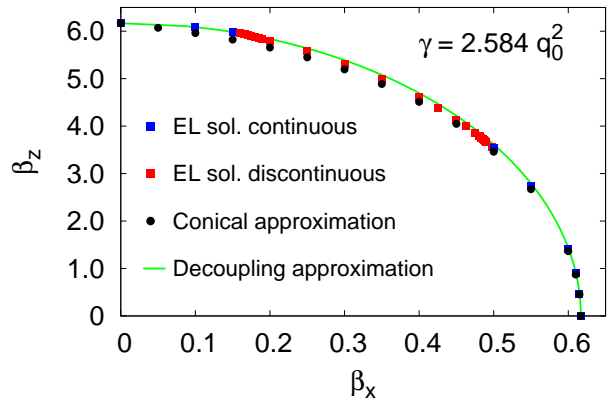}%
\caption{The phase transition line obtained in the conical approximation of 
Ref.~\onlinecite{Chapman14} (black filled circles) and in the decoupling 
approximation of this work (continuous green line) compared with 
the result of the numerical solution of the Euler-Lagrange equations, 
for easy-plane single-ion anisotropy 
$\gamma=0$ (top) and $\gamma=2.584$ (bottom). 
\label{fig:phd_approx}}%
\end{figure}

Minimization of~(\ref{eq:ener_conical}) with respect to $\theta_0$ and
$\kappa$ gives the phase diagram. The IC-FFM transition is everywhere discontinuous except 
at the two
end points of the transition line. Notice that the conical approximation is exact in the
two limiting cases $\beta_x=0$ and $\beta_z=0$. The goodness of the approximation depends
on the magnitude of the easy-plane anisotropy. Fig.~\ref{fig:phd_approx} displays the
phase transition lines obtained in this work by numerical integration of the EL equations
(EL solution), and in the conical approximation, for $\gamma=0$ (top) 
and $\gamma=2.584$ (bottom). In the $\gamma=0$ case the transtion line in the conical approximation 
deviates considerably from the line given by the EL solution. However, for $\gamma=2.584$ the conical 
approximation predicts a transition line which is very close to the EL solution line. 
As the energy of the approximate conical helicoid is always higher than the energy of 
the ground state the approximate transition line lies always below the EL solution transition line, 
as can be appreciated in Fig.~\ref{fig:phd_approx}.

It is easy to understand why the conical approximation locates the transition line very accurately 
as the magnitude of the easy-plane anisotropy increases. 
The energy $\mathcal{H}$ given by Eq.~(\ref{hamil}) contains a 
term $\gamma n_z^2-\beta_z n_z=\gamma(n_z-\beta_z/2\gamma)^2-\beta_z^2/4\gamma$. If $\gamma$ is 
very large the energy penalty caused by a large deviation of $n_z$ from the constant value 
$\beta_z/2\gamma$ cannot be compensated by a gain provided by the other terms in $\mathcal{H}$.
In the vicinity of the transition, however, the energy of the IC state is only slightly lower
than the FFM energy and the conical approximation does not capture the nature of the 
transition.
The differences between the IC and the FFM states are precisely due to the spatial 
variation of the spin. The small variation of $n_z$ allows the IC to reach the FFM state 
continuously. Consider for instance in the transition at constant $\beta_x$ depicted in the 
top left panel of Fig.~\ref{fig:sphere}. The transition takes place continuously when the
spin path (red points) collapse onto the FMM (blue axis). This is not possible in the conical 
approximation, in which the spin configuration paths are forced to follow the parallels of the unit 
spheres of Fig.~\ref{fig:sphere}.
As it gives discontinuous transitions along the whole transition line, the conical approximation 
does not predict the continuous transitions nor the tricritical behaviour.

Let us develop a further approximation, which we call the decoupling approximation, 
in which we assume also a conical ground state. We shall denote by $\beta_{xc}=(\pi^2/16)q_0^2$
and $\beta_{zc}=2\gamma+q_0^2$ the parallel and perpendicular critical fields, respectively. 
We notice that, for large $\gamma$, the value
of $\cos\theta_0$ is determined basically by the three first terms of~(\ref{eq:ener_conical}). The term
proportional to $\beta_x$, which, we recall, is the difference between the energies of the 
conical helicoid~(\ref{eq:sg}) and the conical helix with $\varphi(z)=q_0z$, gives a minor contribution.
Indeed, at low $\beta_x$ and $\beta_z$ the conical helicoid is a slightly distorted conical helix;
at low $\beta_x$ and large $\beta_z$ the term proportional to $\beta_x$ is clearly a small correction;
and for $\beta_x$ close to $\beta_{xc}$ the energy of the conical helicoid is very close to the energy
of the FFM state, which happens to be close to the energy of the conical 
helix\footnote{This is no coincidence, as the phase transition takes place roughly when the energy of 
the helix is similar to the energy of the FFM state.
Indeed, for $\beta_z=0$ we have that the difference between these two energies at the critical
point is $-\beta_{xc}+q_0^2/2=(\pi^2/8-1)q_0^2/2\approx 0.117 q_0^2$.}.
Hence, in the decoupling approximation we have $\cos\theta_0=\beta_z/\beta_{zc}$. 
Minimization of the energy with respect to $\kappa$ gives an equation similar to
that of the well known $\beta_z=0$ case: $E(\kappa)/\kappa=\sqrt{\sin\theta_0\beta_{xc}/\beta_x}$.
The phase transition takes place continuously when $\kappa=1$, that is when
$\beta_x/\beta_{xc}=\sin\theta_0$, and, taking into account the value of $\sin\theta_0$, we get the
following expression for the transition line:
\begin{equation}
\beta_z/\beta_{zc} = \sqrt{1-\beta_x^2/\beta_{xc}^2}\,. \label{eq:TL}
\end{equation}
The continuous green lines of Fig.~\ref{fig:phd_approx} are the curves corresponding to the 
above equation. We see that they describe very accurately the transition line for large easy-plane 
anisotropy, but it departs notably from the EL solution for low $\gamma$. Interestingly, 
equation~(\ref{eq:TL}) shows that the form of the transition line for high anisotropy is universal 
when expressed in terms of the dimensionless fields $\beta_x/\beta_{xc}$ and $\beta_z/\beta_{zc}$. 
We see that the decoupling approximation is as accurate as the conical approximation, or even more, 
and it has the virtue of providing a simple analytic expression for the transition line. On the other 
hand, the decoupling approximation suffers from the same limitations as the conical approximation and 
does not predict correctly the nature of the transition, as it gives always continuous transitions.

\section{Application to \CrNbS \label{sec:exp}}

The general results obtained in this work can be readily applied to \CrNbS. For this compound, 
$a\approx 1.2\,\mathrm{nm}$. Other measurable parameters like the critical parallel and perpendicular
fields, $H_{xc}$ and $H_{zc}$ respectively, and the the zero field helix period, $L_0$, depend strongly 
on the sample. As an instance\cite{Togawa12}, we take $L_0\approx 48\,\mathrm{nm}$ and 
$H_{xc}\approx 2300\,\mathrm{Oe}$. A value $\gamma=2.584q_0^2$ ensures the relation 
$H_{zc}\approx 10 H_{xc}$ tipically observed experimentally. The large easy-plane anisotropy 
implies that the location transition line is accurately given by the results of the conical 
approximation of Ref.~\onlinecite{Chapman14}, and, morover, it is given by the Eq.~(\ref{eq:TL}) 
derived in the present work, so that
\begin{equation}
H_z/H_{zc} = \sqrt{1-H_x^2/H_{xc}^2}\,. \label{eq:phdCr}
\end{equation}
The critical fields $H_{xc}$ and $H_{zc}$ depend strongly on the sample. Due to impurities, 
crystalline defects, etc., each sample will be theoretically described by a different set of the model 
effective parameters.
The anisotropy, however, is expected to be large for all samples, so that Eq.~(\ref{eq:phdCr}) holds 
for any sample, and thus the phase transition line in \CrNbS\ is universal (independent of the sample) 
when expressed in terms of the dimensionless fields $H_x/H_{xc}$ and $H_z/H_{zc}$. 

The position of the tricritical points on the transition line, however, is not universal and depend 
on the sample. For the values reported in Ref.~\onlinecite{Togawa12} ($H_{xc}\approx 2300\,\mathrm{Oe}$ 
and $H_{zc}\approx 23000\,\mathrm{Oe}$) we get $\vec{H}=3729\,\vec{\beta}$~Oe, and thus the tricritical 
points TC1 and TC2 are predicted to be at, respectively, 
$(H_{x}^{t1},H_{z}^{t1})\approx (590, 22200)\,\mathrm{Oe}$, and  
$(H_{x}^{t2},H_{z}^{t2})\approx (1860, 12490)\,\mathrm{Oe}$. 
A CSL can be formed by approaching the CTL2, increasing the perpendicular field $H_x$ 
with the parallel field $H_z$ hold constant below 12500 Oe, or increasing the parallel field 
with the perpendicular field hold constant above 1850 Oe, or increasing a field tilted from
the DM by an angle $\alpha$ smaller than 81 degrees.

Furthermore, a line of discontinuous transitions appears in the phase diagram of \CrNbS, and therefore 
phenomena typical of first order transitions, like phase coexistence and hysteresis in magnetisation and 
magnetoresistance, are expected. 

\section{Conclusions \label{sec:conc}}

In this work the important and long-standing problem of the determination of the phase diagram of 
the mono-axial chiral helimagnet at zero temperature in the magnetic filed plane has been addressed. 
The transition line from the incommensurate phase to the commensurate forced FM state has been thoroghly 
analysed as a function of the single-ion aniotropy and the nature transitions has been elucidated.
We have shown that generically the transition line is composed by a line of discontinuous transition 
separated from two lines of continuous transitions by two tricritical points. 
The nature of the singularities at the
transition line has also been thoroughly analysed and we showed that the transition from the chiral soliton 
lattice to the forced FM state along the continuous transtion line is characterized by logarithmic 
singularities analogous to those characteristic of the case of purely perpendicular magnetic field.
This characteristic singular behaviour may be used to locate one of the tricritical points.

For large anisotropy the conical and the decoupling approximations give the location of the 
transition line very accurately, but they fail in characterising the nature of the transition, 
since the former predicts a discontinuous transition along the whole line and the later
continuous transitions everywhere. Consequently, the tricritical behaviour is not 
predicted by none of these approximations. For large anisotropy we found that the form of the
transition line is universal, given by Eqs.~(\ref{eq:TL}) or~(\ref{eq:phdCr}). The position of the 
tricritical points on the transtion line is not universal, however, and has to be computed for each 
sample.
 
Hence, unexpected predictions for the low temperature regime of \CrNbS\ are given: 
discontinuous transitions, tricritical behaviour, and the universality (that is, the independence of 
the sample) of the form of the transition line due to the large anisotropy characteristic of this 
compound. These phenomena may have interesting applications in spintronics. It remains to study the very 
interesting question of how the phase diagram evolve by increasing the temperature. Work in this 
direction is in progress~\footnote{V. Laliena and J. Campo, in preparation}.

\begin{acknowledgments}
JC and VL acknowledge the grant number MAT2015-68200-C2-2-P.
\end{acknowledgments}

\appendix

\section{Numerical procedure}

The strategy to find a solution to the problem is as follows: for given
$L$ and $\xi_0$ or $\phi_0$, we solve numerically the BVP defined either by~Eqs.~(\ref{eq:EL1}) 
or~(\ref{eq:EL2}); for fixed $L$, we tuned either $\xi_0$ or $\phi_0$ and repeat the computation
until periodicity is reached; then, we compute the energy per period $\mathcal{E}$ via a 
numerical quadrature algorithm. Finally, we determine the period $L$ which minimizes $\mathcal{E}$ 
with a minimisation algorithm. 
Alternatively, we can minimize $\mathcal{E}$ as a function of the two variables $\xi_0$ (or $\phi_0$)
and $L$. Both approaches give the same result, as the minimum of the energy as a function
of the BC for fixed $L$ is the periodic solution which satisfies Eqs.~(\ref{eq:per}).

To solve numerically the BVPs defined by eqs.~(\ref{eq:EL1}) or~(\ref{eq:EL2}) we used a relaxation 
method as described in Ref.~\onlinecite{NRC02} (see also Ref.~\onlinecite{Eggleton71}). The BVP solver 
works as follows. First, the two second order differential equations are converted in a set of
four first order differential equations in the usual way, by introducing two new variables $\omega$
and $v$ and two new equations which relate them to the derivatives of $\varphi$ and $\xi$:
$\omega=d\varphi/dz$, $v=d\xi/dz$. Then, this system of first order differential equations is converted
in a system of finite difference equations by substituting the derivatives by forward finite 
differences. Therefore, a mesh with $N+1$ points $(z_0=0,z_1,\ldots,z_N=L)$ is introduced. 
For simplicity, we used a regular mesh, so that $z_n-z_{n-1}=L/N$ for $n=1,\ldots,N$. 
The finite difference equations and the BC form a set of $4(N+1)$ non-linear algebraic equations 
($4N$ of them given by the difference equations and the remaining 4 provided by the BC) 
with $4(N+1)$ unknowns, which are the values of $\xi(z)$, $\varphi(z)$, $v(z)$, and $\omega(z)$ at the 
mesh points. The algebraic equations are solved numerically with a Newton
algorithm, which proceedes iteratively starting from one initial guess which has to be supplied.
Subsequently, $\mathcal{E}$ is computed with a simple quadrature algorithm using the values $h(z)$ 
calculated at the mesh points.

Periodicity is enforced by tuning $\xi_0$ until the conditions $\omega(L)=\omega(0)$ and $v(L)=v(0)$
are met. The two equations can be enforced by tuning only one variable, $\xi_0$. The reason is that
we exploited translational symmetry to eliminate the freedom in the BC for $\varphi$, which in 
general should read
$\varphi(0)=\varphi_0$ and $\varphi(L)=\varphi_0+2\pi$. Hence, periodicity can be enforced by tuning 
$\varphi_0$ and $\xi_0$. Translational invariance implies that we can set $\varphi_0=0$, taking into 
account that, on physical grounds, there has to be a point in which the spin lies in the plane formed 
by the DM axis and the magnetic field (i.e., with $\varphi=0$). Hence, we enforce the condition 
$v(L)-v(0)=0$ by tuning $\xi_0$ via a simple bracketing-bisection algorithm, and the other condition 
is automatically satisfied. Finally, the minimum of $\mathcal{E}$ as a function of $L$ is obtained 
also with a simple bracketing-bisection algorithm.

To control the numerical errors we set very demanding values for the convergence tolerances of the 
different algorithms: $10^{-14}$ for the BPS solver, $10^{-12}$ for the computation of $\mathcal{E}$,
and $10^{-11}$ for the bracketing-bisection algorithms. The effect of the discretization was taken
into account when computing $\mathcal{E}$ via the quadrature algorithm. The mesh was refined 
iteratively, the number of points being doubled at each iteration, until the difference between the
values of $\mathcal{E}$ computed in two succesive iterations was smaller than the tolerance 
($10^{-12}$). In this way we were able to get the ground state with high accuracy.

The main issue is the convergence of the BVP solver, which works iteratively so that to get convergence 
the initial guess functions have to be close enough to the actual solution. We achieve this by varying 
$\beta_z$ by small steps and using the solution found at one $\beta_z$ as the seed for the next step. 
We start at $\beta_z=0$, taking advantage of the fact that the solution is analytically known there.
This procedure is sound as we expect the IC structure to be continuous on $\vec{\beta}$, as it is in 
the cases where the field is either perpendicular or parallel. The results confirm this point.

\bibliography{references}

\end{document}